\documentclass[twocolumn,showpacs,preprintnumbers,amsmath,amssymb]{revtex4}

\usepackage{graphicx}
\usepackage{dcolumn}
\usepackage{bm}

\begin{document}

\preprint{APS/123-QED}

\title{Spontaneous strain due to ferroquadrupolar ordering in UCu$_2$Sn}

\author{Isao Ishii}
\author{Haruhiro Higaki}%
\author{Toshiro Takabatake}%
\author{Hiroshi Goshima}%
\author{Toshizo Fujita}%
\author{Takashi Suzuki}%
\email{tsuzuki@hiroshima-u.ac.jp}
\affiliation{%
Department of Quantum Matter, ADSM, Hiroshima University, Higashi-Hiroshima 739-8530, Japan\\
}%

\author{Kenichi Katoh}
\affiliation{
Department of Applied Physics, National Defense Academy, Yokosuka 239-8686, Japan\\
}%

\date{\today}

\begin{abstract}
The ternary uranium compound UCu$_2$Sn with a hexagonal ZrPt$_2$Al-type 
structure shows a phase transition at 16 K. 
We reported previously that huge lattice-softening is accompanied by 
the phase transition, which originates from ferroquadrupolar ordering 
of the ground state non-Kramers doublet $\Gamma_5$.
A macroscopic strain, which is expected to emerge spontaneously, 
was not detected by powder X-ray diffraction 
in the temperature range between 4.2 and 300 K. 
To search the spontaneous strain, 
we have carried out thermal expansion measurements on a single-crystalline sample 
along the $a$, $b$ and $c$ axes 
using a capacitance technique with the resolution of $10^{-8}$. 
In the present experiment, 
we found the spontaneous $e_{xx} - e_{yy}$ strain which 
couples to the ground state doublet $\Gamma_5$. 
The effect of uniaxial pressure along the $a$, $b$ and $c$ axes 
on the transition temperature is also discussed. 
\end{abstract}

\maketitle

\section{\label{sec:level1}Introduction
}

Multipolar ordering have been intensively investigated 
in a number of 4$f$-electron compounds~\cite{Morin}. 
In the case of 5$f$-electron systems, however, 
the multipolar ordering has been reported only in a few compounds, 
including NpO$_2$~\cite{Santini2}, 
UPd$_3$~\cite{McEwen}, URu$_2$Si$_2$~\cite{Santini}, UNiSn~\cite{Akazawa} 
and UCu$_2$Sn~\cite{Suzuki}. 
Previously we pointed out that UCu$_2$Sn and UNiSn undergo ferroquadrupolar ordering 
at low temperatures. 

The compound UCu$_2$Sn has a hexagonal ZrPt$_2$Al-type structure (~space group $P6_3/mmc$~) 
with the lattice parameters of $a$~=~4.457~\AA~and $c$~=~8.713~\AA~at 
room temperature. 
In this hexagonal structure, constituent atoms are stacked in layers 
perpendicular to the $c$ axis in regular sequence of $\cdot \cdot \cdot$Sn, Cu, U and 
Cu$\cdot \cdot \cdot$, 
where all U atoms occupy equivalent sites forming a triangle lattice. 
Takabatake $et$~$al$. found that UCu$_2$Sn underwent 
a phase transition around 16 K~\cite{TakabaJPSJ}. 
Afterwords, the transition was estimated to be a non-magnetic one since 
M$\ddot{\rm o}$ssbauer~\cite{Wiese} and NMR~\cite{Kojima} spectroscopies 
inferred the absence of a hyperfine field at Sn and Cu sites 
and neutron diffraction detected no magnetic reflection~\cite{TakabaJMMM}. 
In our previous study on the specific heat and elastic moduli of UCu$_2$Sn~\cite{Suzuki}, 
we determined the crystal electric field ( CEF ) parameters 
( $B_2^0 = 1.682 \times 10$ K, $B_4^0 = -6.100 \times 10^{-2}$ K, 
$B_6^0 = -1.720 \times 10^{-3}$ K, and $B_6^6 = 2.257 \times 10^{-1}$ K ) 
and the CEF level scheme from the ground state non-Kramers doublet $\Gamma_5$ to 
the fifth excited state $\Gamma_3$, 
where $\Gamma_i$ is the irreducible representation for the $6/mmm$ point group. 
We also explained the reasons why the U ions formed the 5$f^2$ configuration with 
the total angular momentum $J = 4$ and the 5$f$-electrons were 
in the localized regime. 
The most prominent feature was that the transverse modulus $C_{66}$ exhibited the huge 
softening around $T_{\rm Q} =$ 16 K, which was an evidence for the ferroquadrupolar 
ordering of the ground state $\Gamma_5$. 
The modulus $C_{66}$ is the linear response to 
$e_{\Gamma_5}$ (~$= e_{xy}$ and 
$= e_{xx} - e_{yy}$~) strain in the hexagonal lattice. 
Taking account of both the strain-quadrupole coupling 
and the quadrupole-quadrupole ( q-q ) coupling, 
we analyzed $C_{66}$ and then obtained 
the positive sign for the q-q coupling coefficient $g'_{\Gamma_5}$, that is, 
ferroquadrupolar coupling in the ground state. 
To distinguish the quadrupolar ordering from the cooperative Jahn-Teller transition, 
we employed a non-dimensional parameter $D \equiv \mid g' C_0 / g^2 N_0 \mid$~\cite{Levy}, 
where $g$ is the strain-quadrupole coupling coefficient, $C_0$ is the background 
value of the elastic modulus and $N_0$ is the number density of 
U ions per unit volume at room temperature. 
The obtained result $D \gg 1$ clearly indicated that the q-q coupling 
$g'$ predominates over the strain-quadrupole coupling $g$ in UCu$_2$Sn and consequently the 
transition is classified as the ferroquadrupolar ordering. 
The ferroquadrupolar ordering must be accompanied by a macroscopic strain or 
distortion below $T_{\rm Q}$. 
In the previous work~\cite{Suzuki} using the powder X-ray diffraction technique, 
we did not succeed in detecting any indication for 
the spontaneous occurrence of macroscopic strain. 
So we made numerical estimation by using the relation 
$| e_{xy} | 
= N_{0}k_{\rm B}g_{\Gamma_{5}}\langle O_{xy} \rangle /C_{0}$~\cite{Luthi} 
with the parameters obtained from fitting the elastic modulus observed, 
and we found that the spontaneous strain might be as small as $5.6 \times 10^{-4}$. 
The value was smaller than the resolution of our X-ray diffraction 
( $\simeq 1 \times 10^{-3}$~). 
In the present work, 
we have carried out the thermal expansion experiments on a single-crystalline sample 
by a capacitance method~\cite{Visser}.

\section{\label{sec:level2}Experimental}

A single crystal of UCu$_2$Sn was grown by a Bridgman method. 
The details of sample preparation was described elsewhere~\cite{TakabaJMMM}. 
An impurity phase of UCuSn (~$\sim$ 4~$\%$ ) was detected in our 
single-crystalline sample of UCu$_2$Sn by the electron probe microanalysis. 
The sample was shaped in a rectangular parallelepiped of 
$2.824 \times 2.908 \times 3.288$ mm$^3$. 
Thermal expansion ${\mit\Delta} l / l$ was measured as a function of temperature $T$ 
from 4.2 to 40 K with a temperature interval of 0.1 K along the $a$, $b$ and $c$ axes 
using a three-terminal method of capacitance measurement. 
Small change in length of the sample was detected by means of change in capacitance 
between the parallel plates with approximately 0.1 mm spacing~\cite{Visser}. 
The plates have an area of $\simeq 1.55 \times 10^2$ mm$^2$. 
The value of $\Delta l / l$ for each axis was defined as 
( $l(T)-l(40{\rm K})$ ) / $l(40{\rm K})$. 
The $a$ and $c$ axes are referred to the international tables 
(~space group $P6_3/mmc$~)~\cite{Hahn}. 
The $b$ axis is defined as perpendicular to the $a$ axis in the hexagonal $c$ plane. 

\section{\label{sec:level3}Results \& Discussion}

\begin{figure}
\includegraphics[height=70mm,clip]{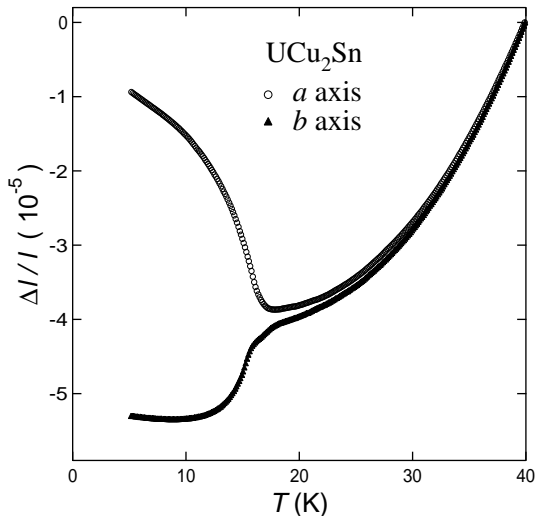}
   \caption{Temperature dependence of thermal expansion $\Delta l / l$ along the $a$ 
and $b$ axes are shown by open circles and solid triangles, respectively.}
   \label{dl_l_ab}
\end{figure}

\begin{figure}
\includegraphics[height=70mm,clip]{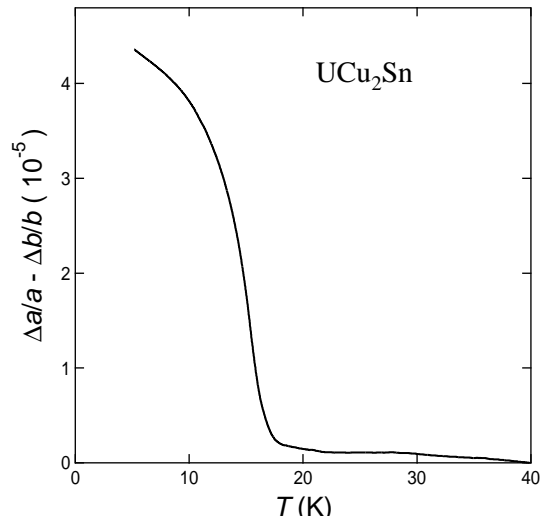}
   \caption{Temperature dependence of $\Delta a / a - \Delta b / b$.}
   \label{strain}
\end{figure}

Figure~\ref{dl_l_ab} shows temperature dependence of thermal expansion 
$\Delta l / l$ both for along the $a$ and $b$ axes. 
At high temperatures, both of $\Delta l / l$ along the $a$ and $b$ axes 
decrease monotonically with decreasing temperature. 
At low temperatures below $T_{\rm Q}$, 
$\Delta l / l$ along the $a$ axis, that is $\Delta a / a$, 
rapidly increases with decreasing temperature, 
whereas $\Delta l / l$ along the $b$ axis, that is $\Delta b / b$, 
continues to decrease. 
As far as the crystal keeps a hexagonal symmetry, 
$\Delta a / a$ and $\Delta b / b$ should coincide with each other 
even though it thermally expands or contracts. 
As clearly seen in Fig.~\ref{dl_l_ab}, $\Delta a / a$ starts to deviate from $\Delta b / b$ 
at a higher temperature than 20 K ( $> T_{\rm Q}$ ). 
This behavior appears to correspond closely to that of the transverse modulus $C_{66}$ 
which starts to soften gradually below $\sim$~20 K. 
The precursor is possibly ascribed to the fluctuation of the quadrupolar ordering. 
Figure \ref{strain} shows the difference $\Delta a / a - \Delta b / b$, 
which is proportional to the expected spontaneous strain $e_{xx} - e_{yy}$. 
Thus, we succeeded in direct confirmation of the macroscopic distortion due to the 
ferroquadrupolar ordering in UCu$_2$Sn. 
The magnitude of the strain evaluated at 5 K is $4.5 \times 10^{-5}$. 
This is the reason why we could not detect any corresponding strain 
by the powder X-ray diffraction with a resolution of $10^{-3}$. 
However, the present value is one order of magnitude smaller than the value of 
$ 5.6 \times 10^{-4}$ which was estimated from the parameter values 
fitted to the elastic modulus observed. 
When a hexagonal system undergoes a structural transition, a 60 degrees 
ferroelastic-type domain is expected to appear in the ordered state. 
In the present case of UCu$_2$Sn, we believe to have observed 
the average of the spontaneous strain over those domains. 
The calculated value of $ 5.6 \times 10^{-4}$ should be regarded as the 
maximum value of the macroscopic strain expected for a single-domain sample. 

The ground state doublet $\Gamma_5$ has a degeneracy of quadrupoles $O_{xy}$ and $O^2_2$. 
One of these order parameters should emerge below $T_{\rm Q}$ and 
therefore the corresponding strain of 
$e_{xy}$ or $e_{xx} - e_{yy}$ 
is expected to appear spontaneously. 
In the present experiment, only the $e_{xx} - e_{yy}$ component 
was detected.
This result strongly suggests that the order parameter is $O_2^2$. 
However, here, we should just notice 
a possibility that the present 
experimental setup may disregard the $e_{xy}$ strain technically even though it emerges. 
\begin{figure}
\includegraphics[height=45mm,clip]{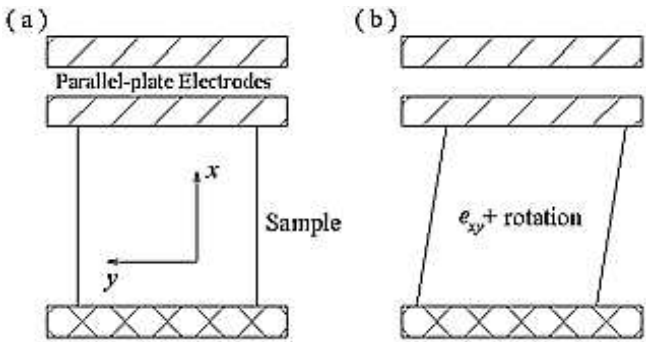}
   \caption{(a) Schematic illustration of the experimental setup for 
the capacitance measurement. 
In this configuration, we can measure the change in the length along the $x$ direction. 
(b) Experimental setup for measuring $e_{xy}$ across $T_{\rm Q}$. 
}
   \label{setup}
\end{figure}
As depicted in Fig.\ref{setup}(a), a change in the sample length along the $x$ direction, 
consequently the strain $e_{xx} - e_{yy}$, 
can be directly measured since we capacitively detect the change in spacing between 
the parallel-plate electrodes. 
In the case of the strain $e_{xy}$, 
the sample will rotate so as to fit the two surfaces of the 
sample onto the parallel plates as shown in Fig.\ref{setup}(b). 
The change $\Delta d$ in the inter-plate spacing will be 
negligibly small because $\Delta d$ is proportional to 
$( 1 - \frac{3}{2} e_{xy}^{2} + \cdot \cdot \cdot )$. 

\begin{figure}
\includegraphics[height=70mm,clip]{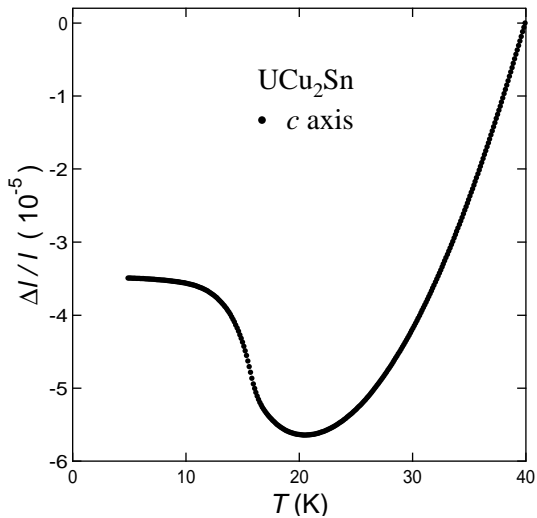}
   \caption{Temperature dependence of thermal expansion $\Delta l / l$ along the $c$ axis.}
   \label{dl_l_c}
\end{figure}

Shown in Fig.~\ref{dl_l_c} is the temperature dependence of thermal expansion 
$\Delta l / l$ along the $c$ axis, that is $\Delta c / c$. 
At high temperatures, $\Delta c / c$ decreases monotonically with 
decreasing temperature. 
It increases gradually below $\sim$~20 K and rapidly below $T_{\rm Q}$. 
We have no convincing explanation for this increase in $\Delta c / c$, 
but a possible origin might be related to development of the secondary order parameter $O_2^0$ 
which couples to $2e_{zz} - e_{xx} - e_{yy}$. 
As we reported previously~\cite{Suzuki}, the strain-quadrupole coupling coefficient 
between $2e_{zz} - e_{xx} - e_{yy}$ and $O_2^0$ is very large. 

\begin{figure}
\includegraphics[height=120mm,clip]{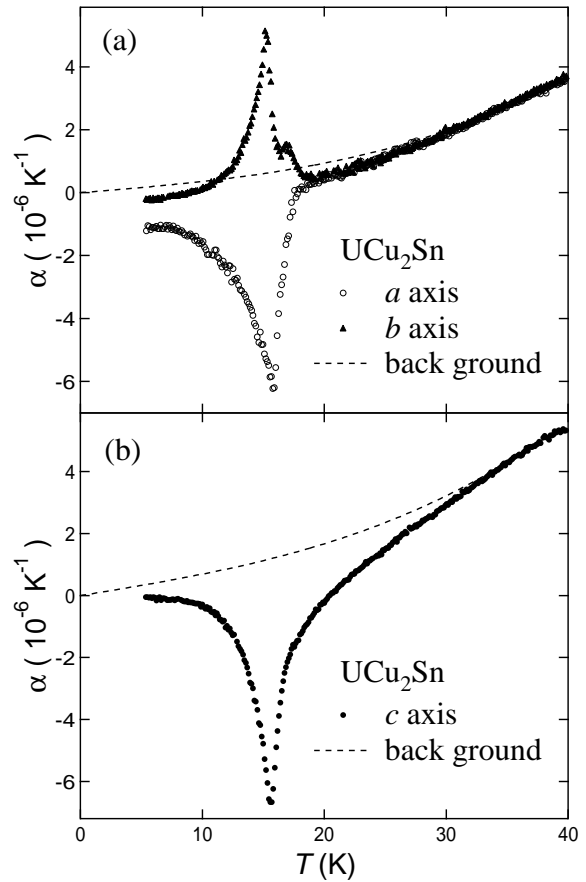}
   \caption{Temperature dependence of the thermal expansion coefficient $\alpha ( T )$. 
(a) Open circles denote $\alpha$ measured along the $a$ axis 
and solid triangles along the $b$ axis. 
The broken curve indicates the background $\alpha_{\rm bg}$. 
(b) Solid circles denotes $\alpha$ along the $c$ axis and 
the broken curve indicates the background.}
   \label{alpha}
\end{figure}

The thermal expansion coefficient $\alpha_i$ is related to $\delta l / l$ 
by the following equation: 
\begin{eqnarray}
\displaystyle \alpha_i = \frac{1}{\delta T}\frac{\delta l_i}{l_i} , \nonumber
\end{eqnarray}
where $\delta$ and the subscript $i$ denote an infinitesimal deference and 
each axis, respectively. 
Figure~\ref{alpha} shows the thermal expansion 
coefficients $\alpha$ as a function of temperature along the $a$, $b$ and $c$ axes. 
Here, we assumed that the background variation of the thermal expansion coefficient is 
given by $\alpha_{\rm bg} = A T + B T^3$~\cite{Barron}. 
The values used for the fitting parameters $A$ and $B$ are listed in Table~\ref{alpha_bg}. 
\begin{table}
\caption{\label{alpha_bg} Fitting parameters $A$ and $B$ for the background $\alpha_{\rm bg}$ 
of thermal expansion coefficients. }
\begin{ruledtabular}
\begin{tabular}{cccc}
axis & $A$ ( K$^{-2}$ ) & $B$ ( K$^{-4}$ ) \\ \hline
$a$, $b$ & $3.21 \times 10^{-8}$ & $3.73 \times 10^{-11}$ \\
$c$ & $6.46 \times 10^{-8}$ & $4.75 \times 10^{-11}$ \\
\end{tabular}
\end{ruledtabular}
\end{table}
From these data, we can estimate the pressure effects on the 
transition temperature $T_{\rm Q}$, 
using the Ehrenfest relation: 
\begin{eqnarray}
\displaystyle \frac{{\rm d} T_{\rm Q}}{{\rm d} P} = 
\frac{\Delta \beta T_{\rm Q} V_{\rm m}}{\Delta C_{\rm p}} , \nonumber
\end{eqnarray}
where the volume expansion coefficient $\Delta \beta$ is assumed as 
$\Delta \beta = \Delta \alpha_a + \Delta \alpha_b +\Delta \alpha_c$. 
$V_{\rm m}$ is the molar volume and 
$\Delta C_{\rm p}$ is the change in the isobaric specific heat at $T_{\rm Q}$. 
We used the difference between $\alpha_{\rm bg}$ and $\alpha_{i}$ for $\Delta \alpha_{i}$ 
at $T_{\rm Q}$. 
The uniaxial pressure effects on the 
transition temperature $T_{\rm Q}$ are estimated from this result. 
The values of ${\rm d} T_{\rm Q} / {\rm d} P_i$ along the $a$, $b$ and $c$ axes 
are listed in Table~\ref{uniaxial_p-dep}. 
\begin{table}
\caption{\label{uniaxial_p-dep} Uniaxial pressure effects on the transition temperature 
$T_{\rm Q}$. The values for ${\rm d} T_{\rm Q} / {\rm d} P_{i}$ are listed in K/GPa. }
\begin{ruledtabular}
\begin{tabular}{cccc}
${\rm d} T_{\rm Q} / {\rm d} P_a$ & ${\rm d} T_{\rm Q} / {\rm d} P_b$ & ${\rm d} 
T_{\rm Q} / {\rm d} P_c$ \\ \hline
$-4.02 \times 10^{-1}$ & $+2.65 \times 10^{-1}$ & $-4.60 \times 10^{-1}$ \\
\end{tabular}
\end{ruledtabular}
\end{table}
To our knowledge, this is the first report on 
the uniaxial pressure effect in UCu$_2$Sn. 
The hydrostatic pressure effect on $T_{\rm Q}$ is also estimated to be 
${\rm d} T_{\rm Q} / {\rm d} P = -6.0 \times 10^{-1}$ K/GPa. 
This value is quite consistent with the value 
${\rm d} T_{\rm Q} / {\rm d} P = -9.6 \times 10^{-1}$ K/GPa 
reported for polycrystalline UCu$_2$Sn by Kurisu $et$~$al$.~\cite{Kurisu}. 

\section{\label{sec:level4}Conclusion}

We measured the thermal expansion along the $a$, $b$ and $c$ axes 
of single-crystalline UCu$_2$Sn. 
The change in the thermal expansion below $T_{\rm Q}$ 
clearly indicates the spontaneous emergence of the macroscopic strain 
$e_{xx} - e_{yy}$, 
which couples to the quadrupole $O_2^2$. 
As a result, it is completely proved that the transition in UCu$_2$Sn at $T_{\rm Q}$ 
originates from the ferroquadrupolar ordering. 
The enhancement of $\Delta c / c$ below $T_{\rm Q}$ might be regarded as due to 
the development of the secondary order parameter $O_2^0$. 
We also discussed the uniaxial pressure effect on $T_{\rm Q}$, 
and succeeded in evaluating ${\rm d} T_{\rm Q} / {\rm d} P_i$. 

\section{\label{sec:level5}Acknowledgments}

This work was supported by Grant-in-Aids for both 
Scientific Research (B) (No.13440114) and 
COE Research (No.13CE2002) from the Ministry of 
Education, Culture, Sports, Science and Technology of Japan. 
We thank the Cryogenic Center of Hiroshima University for their experimental backup. 



\begin{references}
%
\bibitem{Morin} P.Morin, D.SchmittFin Ferromagnetic Materials, vol.5 (North-Holland, 
Amsterdam 1990) ed. K.H.J.Buschow and E.P.Wohlfarth, p.1. 
%
\bibitem{Santini2} P.Santini and G.Amoretti Phys. Rev. Lett. {\bf 85}, 2188 (2000). 
%
\bibitem{McEwen} K.A.McEwen, U.Steigenberger, K.N.Clausen, J.Kulda, J.G.Park, and M.B.Walker
 J. Magn. Magn. Mater. {\bf 177-181}, 37 (1998). 
%
\bibitem{Santini} P.Santini and G.Amoretti Phys. Rev. Lett. {\bf 73}, 1027 (1994). 
%
\bibitem{Akazawa} T.Akazawa, T.Suzuki, H.Goshima, T.Tahara, T.Fujita, T.Takabatake 
and H.Fujii J. Phys. Soc. Jpn. {\bf 67}, 3256 (1998). 
%
\bibitem{Suzuki} T.Suzuki, I.Ishii, N.Okuda, K.Katoh T.Takabatake, T.Fujita and A.Tamaki 
Phys. Rev. B {\bf 62}, 49 (2000). 
%
\bibitem{TakabaJPSJ} T.Takabatake, H.Iwasaki, H.Fujii, S.Ikeda S.Nishigori, 
Y.Aoki, T.Suzuki and T.Fujita 
J. Phys. Soc. Jpn. {\bf 61}, 778 (1992). 
%
\bibitem{Wiese} S.Wiese, E.Gamper, H.Winkelmann, B.B$\ddot{\rm u}$chner, M.M.Abd-Elmeguid, 
H.Micklitz, T.Takabatake 
Physica B {\bf 230-232}, 95 (1997). 
%
\bibitem{Kojima} K.Kojima, A.Harada, T.Takabatake, S.Ogura, K.Hiraoka, 
Physica B {\bf 269}, 249 (1999). 
%
\bibitem{TakabaJMMM} T.Takabatake, M.Shirase, K.Katoh, Y.Echizen, K.Sugiyama and T.Osakabe 
J. Magn. Magn. Mater. {\bf 177-181}, 53 (1998). 
%
\bibitem{Levy} P.M.Levy, P.Morin, and D.Schmitt Phys. Rev. Lett. {\bf 42}, 1417 (1979). 
%
\bibitem{Luthi} B. L\"{u}thi, in {\it Dynamical Properties of Solids} edited 
G. K. Horton and A. A. Maradudin (NorthHolland, Amsterdam 1980). 
%
\bibitem{Visser} A.de Visser :Ph.D.Thesis, Amsterdam university (1986). 
%
\bibitem{Hahn} T.Hahn (ed.) "International Tables for Crystallography, 
Brief teaching edition of volume A, Space-group symmetry 2nd, rev. ed" 
Kluwer Academic Publishers, (1989). 
%
\bibitem{Barron} T.H.K.Barron, J.G.Collons and G.K.White 
Advances Physics {\bf 29}, 609 (1980). 
%
\bibitem{Kurisu} M.Kurisu, T.Takabatake, H.Iwasaki and H.Fujii 
Physica B {\bf 206 $\&$ 207}, 505 (1995). 
%
\end{references}
\end{document}